\documentclass[aps,prl,preprint,superscriptaddress]{revtex4}
\usepackage{graphicx}
\usepackage{color}
\usepackage{amsfonts}
\usepackage{ulem}

\definecolor{brown}{RGB}{200,100,0}

\def\kbar{$\bar{\mathrm{K}}$}
\def\mbar{$\bar{\mathrm{M}}$}
\def\kbarp{\kbar$^{\prime}$}
\def\gbar{$\bar{\mathrm{\Gamma}}$}
\def\kgk{\kbar-\gbar-\kbarp}
\def\WS2{WS$_2$}
\def\MoS2{MoS$_2$}{

\begin{document}
\title{Nanoscale patterning of quasiparticle band alignment}
\author{ S\o ren Ulstrup}
\email{ulstrup@phys.au.dk}
 \affiliation{Department of Physics and Astronomy, Interdisciplinary Nanoscience Center, Aarhus University,
8000 Aarhus C, Denmark}
\author{Cristina E. Giusca}
\email{cristina.giusca@npl.co.uk}
\affiliation{National Physical Laboratory, Hampton Road, Teddington TW11 0LW, United Kingdom } 
\author{Jill A. Miwa}
 \affiliation{Department of Physics and Astronomy, Interdisciplinary Nanoscience Center, Aarhus University,
8000 Aarhus C, Denmark}
\author{Charlotte E. Sanders}
\affiliation{Central Laser Facility, STFC Rutherford Appleton Laboratory, Harwell, United Kingdom}
\author{Alex Browning}
\affiliation{National Physical Laboratory, Hampton Road, Teddington TW11 0LW, United Kingdom } 
\author{Pavel Dudin}
\author{Cephise Cacho}
\affiliation{Diamond Light Source, Division of Science, Didcot, United Kingdom}
\author{Olga Kazakova}
\affiliation{National Physical Laboratory, Hampton Road, Teddington TW11 0LW, United Kingdom } 
\author{ D. Kurt Gaskill}
\author{Rachael L. Myers-Ward}
\affiliation{U.S. Naval Research Laboratory, Washington, DC 20375, United States of America}
\author{Tianyi Zhang}
\author{Mauricio Terrones}
\affiliation{Department of Physics and Center for 2-Dimensional and Layered Materials, Department of Materials Sciences and Engineering, The Pennsylvania State University, Pennsylvania 16802, United States of America}
\author{ Philip Hofmann}
\affiliation{Department of Physics and Astronomy, Interdisciplinary Nanoscience Center, Aarhus University,
8000 Aarhus C, Denmark}
\maketitle

\textbf{Control of atomic-scale interfaces between materials with distinct electronic structures is crucial for the design and fabrication of most electronic devices. In the case of two-dimensional (2D) materials, disparate electronic structures can be realized even within a single uniform sheet, merely by locally applying different vertical bias voltages \cite{Baugher:2014}. Indeed, it has been suggested that nanoscale electronic patterning in a single sheet can be achieved by placing the 2D material on a suitably pre-patterned substrate \cite{Rosner:2016aa}, exploiting the sensitivity of 2D materials to their environment via  band alignment \cite{Lee:2014}, screening \cite{Komsa:2012aa,Ugeda:2014,Antonija-Grubisic-Cabo:2015aa} or  hybridization \cite{Allain2015,Dendzik:2017,Shao:2019aa}.  Here, we utilize the inherently nano-structured single layer (SL) and bilayer (BL) graphene on silicon carbide to laterally tune the electrostatic gating of adjacent SL tungsten disulphide (WS$_2$) in a van der Waals heterostructure. The electronic band alignments are mapped in energy and momentum space using angle-resolved photoemission with a spatial resolution on the order of 500~nm  (nanoARPES). We find that the SL WS$_2$ band offsets track the work function of the underlying SL and BL graphene, and we relate such changes to observed lateral patterns of exciton and trion luminescence from SL WS$_2$, demonstrating ultimate control of optoelectronic properties at the nanoscale.}

The construction of a 2D electronic device, such as a $pn$-junction can be envisioned using two strategies: The first is to smoothly join two 2D materials with different electronic properties, essentially following the established recipe for 3D semiconductors. Alternatively, one can exploit the 2D materials' sensitivity to their environment and create junctions using a single uniform sheet of material placed over a suitably pre-patterned substrate \cite{Rosner:2016aa,Wilson:2017aa,Huang:2018ab}. This approach has several advantages, such as technical simplicity and the absence of a possibly defective interface \cite{Wang:2013,Zhang:2018}. However, the interaction between a 2D material and substrate is highly non-trivial and hitherto poorly understood: Even in the absence of hybridization or charge transfer, substrate-screening can lead to an asymmetric band gap change, creating a type II heterojunction within a single sheet of 2D material \cite{Rosner:2016aa}. Moreover, strong many-body effects lead to a complex connection between the quasiparticle band structure and the optical properties. On one hand, even strong changes of the quasiparticle band structure might only have a very minor influence on the optical band gap, due to the interplay of the quasiparticle band gap size and exciton binding energy \cite{Komsa:2012aa}. On the other hand, the quasiparticle band structure can greatly affect the formation of more complex entities such as trions \cite{Katoch:2018}. 

\begin{figure*}
\begin{center}
\includegraphics[width=0.98\textwidth]{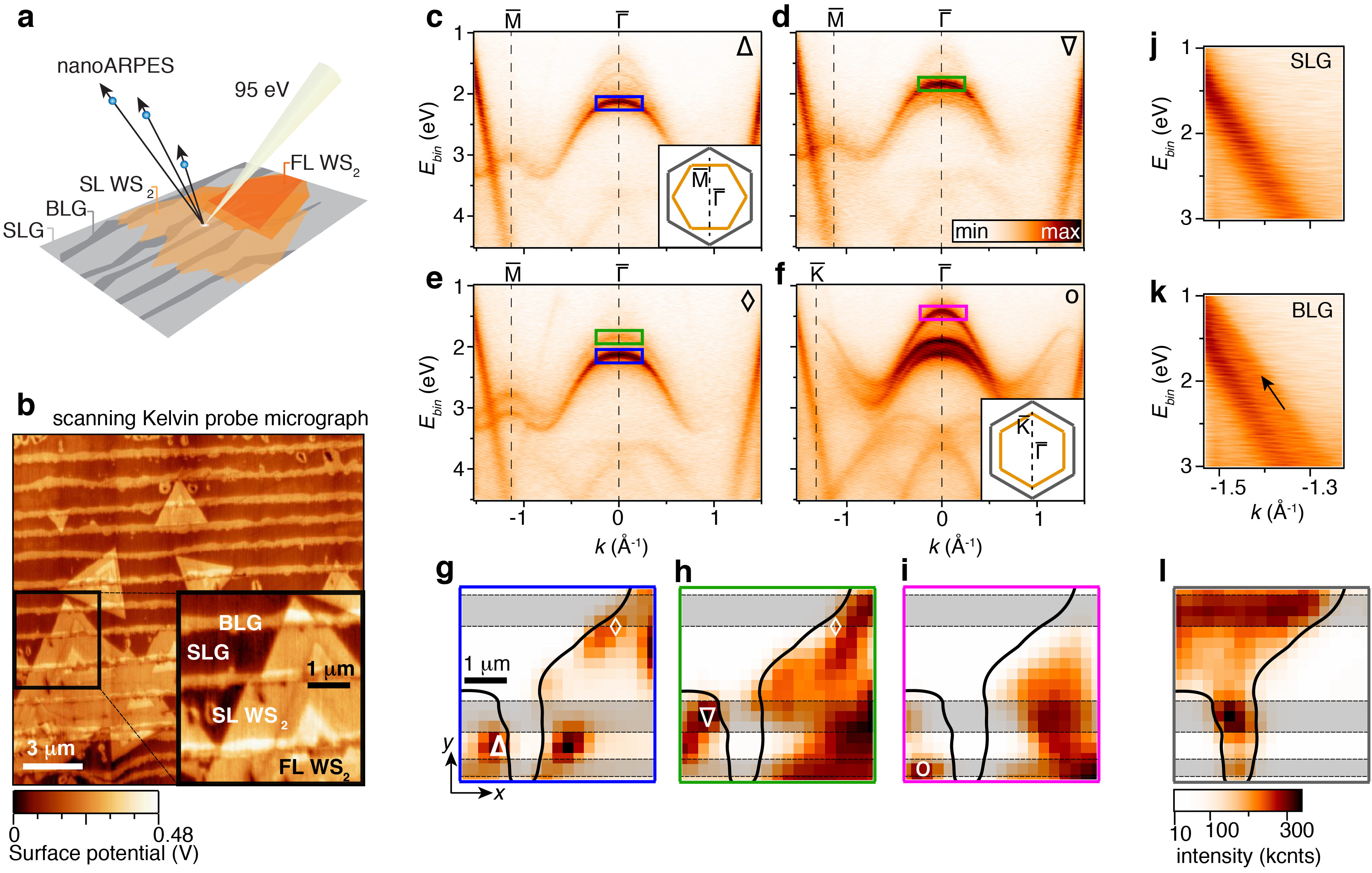}
\caption{\textbf{Nanoscale spatial mapping of WS$_2$ electronic dispersion on a nano-patterned graphene substrate.} \textbf{a,} Sketch of the nanoARPES experiment. \textbf{b,} Scanning Kelvin probe micrograph showing the variation in surface potential over a typical area of the sample. The inset highlights an area containing SLG and BLG, as well as, SL and FL WS$_2$, similar to the samples measured with nanoARPES. \textbf{c-f,} Representative ARPES spectra binned over ($500 \times 500$)~nm$^2$ areas in the real space maps of the photoemission intensity in panels \textbf{g-i}. The symbols $\bigtriangleup$, $\bigtriangledown$, $\diamond$, $\circ$ on the maps in \textbf{g}-\textbf{i} indicate where a given $(E,k)$-dispersion in the correspondingly labeled panel in \textbf{c}-\textbf{f} was extracted. The Brillouin zones in the insets of \textbf{c} and \textbf{f} (grey for graphene, orange for WS$_2$) give the orientation of the WS$_2$ island in the acquisition area with respect to the underlying graphene for panels \textbf{c}-\textbf{e} and \textbf{f}, respectively, along with the cut direction for the measurement (black dashed line). High symmetry points refer to the  WS$_2$ Brillouin zone. All data have been acquired along the \kgk~high symmetry direction of the single-domain graphene Brillouin zone as explained in Supplementary Section 1. The intensity maps in \textbf{g-i} are composed from the photoemission intensity in the $(E,k)$-regions marked with boxes of the same color in \textbf{c}-\textbf{f}. An outline of the WS$_2$ island edges has been drawn and the underlying BLG stripes are highlighted by grey-shaded boxes. \textbf{j-k,} Detailed dispersion around the graphene branches from \textbf{j} SLG and \textbf{k} BLG stripes. \textbf{l,} Intensity map obtained from the BLG branch marked by an arrow in \textbf{k}. The color scale bar in \textbf{l} also applies to \textbf{g-i}.}
\label{fig:1}
\end{center}
\end{figure*}

Here we investigate the interplay of quasiparticle band alignments and optical properties in a lateral heterostructure of semiconducting SL WS$_2$ placed on alternating nanostripes of SL graphene (SLG) and BL graphene (BLG) grown on SiC. This strategy enables us to pattern the electronic structure and light-matter interaction at the nanoscale due to the laterally varying work functions between SLG and BLG \cite{Giusca:2016,Kastl:2019}. We directly visualize how the electronic structure changes at the complex heterogeneous atomic-scale interfaces present in our samples using nanoARPES; see illustration in Fig. \ref{fig:1}(a). This groundbreaking technique for electronic structure characterization provides three key new insights for the type of van der Waals heterostructure investigated here: (i) We can determine the energy- and momentum-dependence of band alignments at truly 2D interfaces, (ii) we obtain detailed spatially resolved information on how the electronic structure of a 2D semiconductor is modified around the one-dimensional (1D) SLG/BLG interface, and (iii) we can spatially disentangle the electronic dispersions of SL WS$_2$ and few-layer (FL) WS$_2$, and distinguish between islands of different orientations.  

%% From here we can follow the existing flow.
Fig. \ref{fig:1}(b) shows the morphology and microscopic surface potential of  WS$_2$ islands on graphene measured by scanning Kelvin probe microscopy (SKPM) under ambient conditions. Triangular WS$_2$ islands are observed with SL regions near the edges and FL areas towards the center.  Alternating stripes of BLG and SLG are visible in both bare and WS$_2$-covered areas. The strong contrast difference between WS$_2$ placed on alternating stripes of BLG and SLG is caused by the large work function difference on the order of 100~meV \cite{Giusca:2015}. The SL WS$_2$ islands have a negligible influence on the relative work function difference between the underlying SLG and BLG, as confirmed by density functional theory calculations \citep{Giusca:2015}.

Figures \ref{fig:1}(c)-(f) present the $(E,k)$-dependence of the topmost WS$_2$ valence bands (VBs) measured from (500$\times$500)~nm$^2$ areas on the sample using nanoARPES, extracted at the locations indicated with corresponding markers on the real space maps in Fig. \ref{fig:1}(g)-(i). Typically, a sharp and intense state is observed at \gbar~that can be assigned to the local VB maximum (VBM) of SL WS$_2$ \cite{Henck:2018,Kastl:2018}. Upon close inspection, the binding energy of the VBM turns out to depend on the position within a WS$_2$ island.  In most cases, the VBM is found at either  the energy shown in  Fig. \ref{fig:1}(c) or that in Fig. \ref{fig:1}(d). These two different energy regions have thus been marked by a blue and green box, respectively.  In some areas, it is even possible to observe the simultaneous presence of two rigidly shifted SL WS$_2$ VBs (Fig. \ref{fig:1}(e)). The dispersion in Fig. \ref{fig:1}(f), on the other hand, is strikingly different from the other examples, showing a three-fold splitting with nearly equal intensity distribution between the split bands at \gbar. The WS$_2$ islands tend to orient either with the \gbar-\mbar~(see Figs. \ref{fig:1}(c)-(e)) or the \gbar-\kbar~(see Fig. \ref{fig:1}(f)) high symmetry directions aligned with the underlying graphene, although we occasionally find other orientations. 

%Upon close examination of the VB binding energy positions, an offset between the two dispersions is seen. A similar and more clear example of this is obtained in a different area in Fig. \ref{fig:1}(e) where a superposition of two rigidly shifted SL WS$_2$ VBs is visible along \gbar-\mbar. The dispersion in Fig. \ref{fig:1}(f) is strikingly different from the other examples as a three-fold splitting with nearly equal intensity distribution between the split bands occurs at \gbar. The WS$_2$ islands tend to orient either with the \gbar-\mbar~(see Figs. \ref{fig:1}(c)-(e)) or the \gbar-\kbar~(see Fig. \ref{fig:1}(f)) high symmetry directions aligned with the underlying graphene, although we occasionally find other orientations. 

Further insight into local variations in the dispersion is obtained by investigating the spatial intensity distribution of the split states at \gbar, as shown in Figs. \ref{fig:1}(g)-(i). These images correspond to real space maps of the photoemission intensity composed from the $(E,k)$-regions demarcated by boxes of the same color in Figs. \ref{fig:1}(c)-(f). The maps have been measured in scanning steps of 250~nm over a (4.5$\times$4.5 $)\mu$m$^2$ area, thereby covering the edges of two adjacent WS$_2$ islands of different orientations, as in the very similar region imaged by SKPM in the inset of Fig. \ref{fig:1}(b).  The two SL WS$_2$ VBs at different binding energy positions originate from distinct areas close to the edges where they give rise to the intense spots in Figs. \ref{fig:1}(g)-(h). The topmost split VB states (see magenta box in Fig. \ref{fig:1}(f)) are concentrated towards the interior of the WS$_2$ islands, where mainly FL structures occur, as evidenced by SKPM in Fig. \ref{fig:1}(b). In fact, the band structure in Fig. \ref{fig:1}(f) is easily identified as being caused by multilayer splitting rather than simple shifts due to the visibly different effective mass (inverse curvature) of the topmost band. 

We show that the shift between the valence bands in Figs. \ref{fig:1}(c)-(e) is correlated with the thickness of the underlying graphene by composing a real space map from the photoemission intensity of a BLG band. BLG is characterized by a splitting of the linear $\pi$-band near the \kbar~point as shown in Figs. \ref{fig:1}(j)-(k) (see arrow in panel~(k) for the second branch). Mapping the intensity from this second branch permits a straightforward identification of BLG stripes in Fig. \ref{fig:1}(l); and this has been used to mark the grey-shaded boxes in all the real space maps. The BLG stripes are found to coincide with areas where the SL WS$_2$ VB is shifted to lower binding energies, see Fig. \ref{fig:1}(d) and (h). Additional details of the correlation between graphene thickness and SL WS$_2$ VB binding energy positions are discussed in Supplementary Section 3 and 4.

\begin{figure*}
\begin{center}
\includegraphics[width=0.8\textwidth]{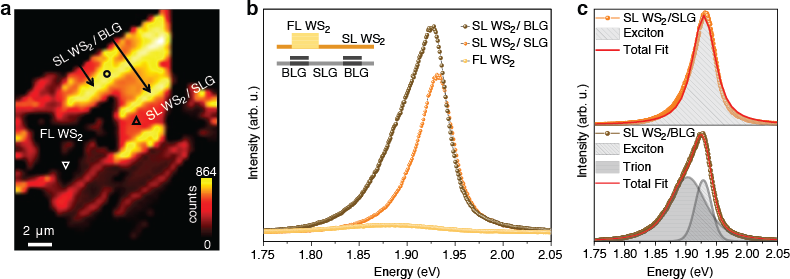}
\caption{\textbf{Spatial distribution of photoluminescence intensity and associated spectral response.} \textbf{a,} Photoluminescence (PL) intensity map of representative WS$_2$  island showing alternating regions of enhanced intensity and quenched signal for thick area at the centre of the island. The enhanced PL coincides with areas where WS$_2$ overlaps with BLG stripes, with some examples (yellow stripes) indicated by black arrows. The wider areas of lower PL intensity (red colour) correspond to SL WS$_2$ on SLG. \textbf{b,} Comparison of photoluminescence spectra associated with SL WS$_2$ on SLG, SL WS$_2$ on BLG and with FL WS$_2$. These are extracted from the locations indicated by symbols in \textbf{a}. \textbf{c,} PL peak deconvolution, carried out using Lorentzian shape components, for SL WS$_2$ on SLG (top panel) and SL WS$_2$ on BLG, highlighting the appearance of the trion (bottom panel). 
}
\label{fig:2}
\end{center}
\end{figure*}

% Maybe point out that the role of the substrate has been previously assigned to t
We turn to the consequences of this spatially heterogeneous electronic structure for the luminescence of optically excited electron-hole pairs, i.e. excitons, and charged excitons (trions) in WS$_2$ \cite{Mak:2016}. Photoluminescence (PL) mapping of a WS$_2$ island, acquired under ambient conditions, is shown in Fig.  \ref{fig:2}(a), where a stronger PL signal is observed on SL WS$_2$ on BLG compared to SL WS$_2$ on SLG. 
The energies of characteristic lines associated with SL WS$_2$ on SLG and on BLG are identified in the PL spectra displayed in Fig. \ref{fig:2}(b), which also shows the weak PL response for the FL WS$_2$ area at the centre of the island, which is mainly due to the indirect-band gap character in the bulk. Detailed analysis by curve fitting to Lorentzian line shapes in Fig. \ref{fig:2}(c) reveals an additional component for SL WS$_2$ on BLG (bottom panel), attributed to charged exciton states (trions) at an energy of 1.90 eV, whereas the neutral exciton peak is found at 1.93 eV for both WS$_2$ on SLG (top panel) and BLG.

\begin{figure*}
\begin{center}
\includegraphics[width=0.98\textwidth]{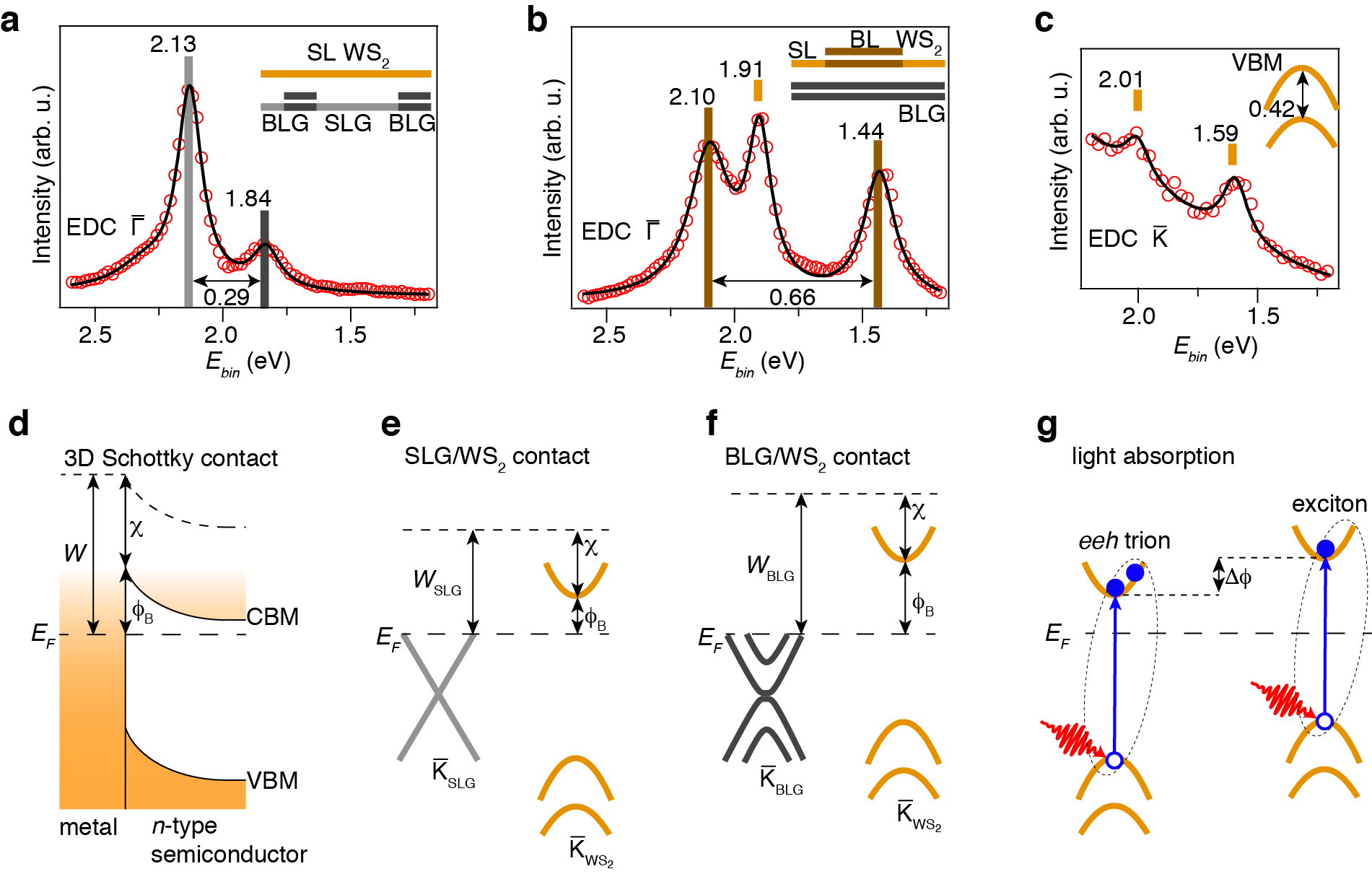}
\caption{\textbf{Analysis of band alignments } \textbf{a-b,} Energy distribution curves (EDCs, red circles) binned within $\pm$0.1~\AA$^{-1}$ around the \gbar-point in the dispersions in Figs. 1(e)-(f), respectively. The analysis is carried out \textbf{a} on a SL WS$_2$ island straddling SLG and BLG patches, and \textbf{b} on mixed SL and BL WS$_2$ supported on BLG (see side views in the insets). Peak positions obtained from fits to Lorentzian line shapes (black curves) on a constant background are given in units of electron volts. The tick marks above the peaks have been coloured according to the structural diagrams and the spatial region where the peak originates from. \textbf{c}, Similar EDC analysis as in \textbf{a-b} but carried out at the \kbar-point in the dispersion in Fig. 1(f). The fitted peak positions provide the VBM and spin-orbit splitting (see inset). \textbf{d}, Typical band diagram for a 3D Schottky contact between a metal and an $n$-type semiconductor. The double-headed arrows illustrate the alignments of work function $W$, Schottky barrier height $\phi_B$ and electron affinity $\chi$. \textbf{e-f}, 2D Schottky alignment diagrams for \textbf{e} a SLG/WS$_2$ contact and \textbf{f} a BLG/WS$_2$ contact in UHV. The trend of the band alignments has been derived from the EDCs in \textbf{a-c}.  \textbf{g}, Sketch of optically induced trion (electron-electron-hole ($eeh$) type) and exciton processes in SL WS$_2$ under a different bias $\Delta \phi$ caused by the varying Schottky barrier heights. Full (hollow) spheres correspond to electrons (holes), the red arrow corresponds to an optical pulse and the blue arrow signifies an electron-hole pair excitation.}
\label{fig:3}
\end{center}
\end{figure*}

We have now tracked both the band offsets and the excitonic spectrum across the SLG-BLG interface beneath SL WS$_2$ and are thus in a position to explore the connection between these. In order to obtain more accurate values for the band offsets, we analyze energy distribution curve (EDC) cuts at \gbar~for the different structures. Fig. \ref{fig:3}(a) presents an EDC from the spectrum in Fig. \ref{fig:1}(e) where a SL WS$_2$ island straddles SLG and BLG stripes (see inset in Fig. \ref{fig:3}(a)). Curve fitting of the peak positions reveals a binding energy shift of the WS$_2$ of 0.29(5)~eV between SLG and BLG. Performing a similar EDC analysis of the spectrum in Fig. \ref{fig:1}(f) reveals that a splitting of 0.66(1)~eV occurs between the states at lowest and highest binding energies, which matches the expected splitting of BL WS$_2$ \cite{Zeng:2013}. The additional peak at 1.91(2)~eV between the BL WS$_2$ bands is attributed to a SL region (see inset in Fig. \ref{fig:3}(b)) on a BLG stripe. 

The data in Fig. \ref{fig:1}(f) also provides access to the \kbar~point~of WS$_2$, which is characterized by spin-orbit split bands and forms the global VBM in SL WS$_2$. \kbar~is not accessible in the other spectra in Fig. \ref{fig:1}(c)-(e) because of the rotated BZ. The EDC fit in Fig. \ref{fig:3}(c) yields a spin-orbit splitting of 0.42(4)~eV, in agreement with previous studies of SL WS$_2$ in van der Waals heterostructures \cite{Katoch:2018,Kastl:2018}, and a VBM of 1.59(4)~eV for SL WS$_2$ on BLG. By rigidly correcting for the shift on SLG areas one would thus expect the VBM on those regions around a binding energy of 1.9~eV. Under the assumption that the direct quasiparticle band gap of SL WS$_2$ on SLG and BLG is smaller than 2.4 eV measured on silica \cite{Chernikov:2015b}, we can infer that our WS$_2$ remains $n$-type doped in the entire sample, although the density of free electrons will be substantially higher on SLG. 

The rigid VB shifts of WS$_2$ on SLG and BLG areas are consistent with an ideal 2D Schottky contact between WS$_2$ and graphene. In order to see this, consider first a sketch of the band alignments for 3D metal-semiconductor junctions in Fig. \ref{fig:3}(d). The Schottky barrier height $\phi_B$ is set by the metal work function $W$ and semiconductor electron affinity $\chi$, i.e. $\phi_B = W - \chi$. Forming a metal-semiconductor contact  leads to band bending with a depletion region towards the bulk of the semiconductor. For the interface between two 2D materials, this is irrelevant and the band offset is expected to follow the sketch in Figs. \ref{fig:3}(e)-(f) for WS$_2$ on SLG and BLG, respectively.  In this case, the band offset is expected to follow the work function change between SLG and BLG on SiC, such that the higher work function of BLG pushes the WS$_2$ VBM closer to $E_F$ \cite{Mammadov:2017}, as observed in our data. The difference in Schottky barrier height between SLG and BLG areas results in a built-in bias $\Delta \phi \approx 0.3$~eV that laterally conforms to the SLG/BLG patterns. The magnitude of $\Delta \phi$ is similar to the SLG/BLG work function difference of 0.1--0.2 eV in ultra-high vacuum (UHV) \cite{Mammadov:2017}. We speculate that the slightly larger shift in our case can be attributed to a difference in dielectric screening between SLG and BLG, which may give rise to an asymmetric renormalization of the WS$_2$ quasiparticle gap, effectively causing a variation in $\chi$ as well \cite{Ugeda:2014,Rosner:2016aa}. The interpretation of the band alignment in terms of a Schottky contact without Fermi level pinning relies on the quasi-freestanding nature of WS$_2$ on graphene \cite{Allain2015,Quang:2017}. It is  consistent with the absence of  hybridization between graphene and WS$_2$ bands in any of our spectra, as well as with the sharp VB features at \gbar, in contrast to the situation on metal substrates \cite{Dendzik:2015,Dendzik:2017}.

We conclude by interpreting the nano-patterned PL signal in Fig. \ref{fig:2} in terms of the Schottky contact-induced band alignment. Superficially, it appears surprising that the exciton PL energy is nearly identical for the two sample regions with  different band alignments. This, however, is well understood. A rigid band offset would not be expected to affect the quasiparticle band gap in the material, and even a screening-induced band gap renormalization, would only be expected to have a minor effect on the exciton binding energy \cite{Komsa:2012aa}. The change in band alignment can be used to explain the strongly increased trion signal in the BLG areas, as indicated in Fig. \ref{fig:3}(g). The more $n$-doped WS$_2$ would have a strongly increased population of electrons in the conduction band, facilitating the formation of negatively charged electron-electron-hole ($eeh$) trions when the material is excited by light as sketched in Fig. \ref{fig:3}(g) \cite{Mak:2013,Katoch:2018}. Our nanoARPES measurements suggest that trion formation would be expected in the SLG areas, whereas our PL measurements indicate that it is actually favoured in the BLG areas (Fig. \ref{fig:2}). This can still be understood in terms of Fig. \ref{fig:3}(g), combined with the knowledge that the PL maps were acquired under ambient conditions rather than in UHV. Under ambient conditions,  the higher reactivity of SLG compared to BLG leads to the adsorption of impurities and a \textit{reversal} of the work function difference in the SLG/BLG patterns compared to UHV, as explained in more detail in Supplementary Section 5 \cite{Giusca:2015}. Since the deposited WS$_2$ largely tracks the work function of the underlying SLG/BLG \cite{Giusca:2016}, this is accompanied by a reversal of the band alignment. We note that the band offset between the SLG and BLG regions also implies the existence of a 1D interface with lateral band bending in the WS$_2$ VBs, but this is not observable in our experiments because the screening length of graphene on SiC is an order of magnitude smaller than our spatial resolution, as discussed in Supplementary Section 6. 

The sharp 1D interfaces and the laterally varying gating of WS$_2$ between SLG/BLG stripes demonstrates the concept of creating nanoscaled devices from a single sheet of 2D material, placed on a suitably patterned substrate. Particularly intriguing is the complex interplay between electronic and optical properties, that not only allows the confinement of electronic states but also that of more complex objects -- such as trions -- on the nanoscale, opening a promising avenue for engineering novel 2D devices. 

\section{Methods}
\textbf{Growth of WS$_2$/Graphene/SiC heterostructures.} Graphene was synthesised on a semi-insulating (0001) 6H-SiC substrate etched in H$_2$ at 200 mbar, during a temperature ramp from room temperature to 1580 $^{\circ}$C to remove polishing damage. Graphene growth was carried out at 1580 $^{\circ}$C for 25 minutes in Ar gas, at 100 mbar, and used as a substrate for subsequent WS$_2$ growth. WS$_2$ islands were synthesized on graphene/SiC at 900$^{\circ}$C by ambient pressure chemical vapour deposition. During the synthesis process, sulphur powders were heated up to 250$^{\circ}$C to generate sulphur vapour. Ar gas flow was used for carrying the sulphur vapour to react with WO$_3$ powder.

\textbf{Scanning Kelvin probe microscopy.} SKPM experiments were carried out in ambient conditions, using a Bruker Icon AFM and Bruker highly doped Si probes (PFQNE-AL) with a force constant $\approx$0.9 N/m and resonant frequency $f_0$ of 300 kHz. Double-pass frequency-modulated SKPM (FM-SKPM) has been used in all measurements, with topography acquired first and the surface potential recorded in a second pass. An AC voltage with a lower frequency ($f_{mod}$ = 3 kHz) than that of the resonant frequency of the cantilever was applied to the tip, inducing a frequency shift. The feedback loop of FM-KPFM monitored the side modes, $f_0 \pm f_{mod}$, and compensated the mode frequency by applying an offset DC voltage, equal to the contact potential difference, which was recorded to obtain the surface potential map.
The FM-SKPM experiments in Supplementary Section 5 were carried out in ambient air and vacuum (pressure of $1 \times 10^{−6}$ mbar) as described above using an NT-MDT NTEGRA Aura system.\\

\textbf{Photoluminescence mapping.} PL spectroscopy mapping was carried out under ambient conditions using a Renishaw inVia confocal microscope-based system using a 532 nm laser line as the excitation wavelength (2.33 eV excitation energy). The laser beam was focused through a 100$\times$ microscope objective, with the PL signal recorded in back-scattering geometry, using integration time of 0.1 s/pixel and a lateral spacing of 0.3 $\mu$m to acquire the PL intensity maps.\\

\textbf{nanoARPES.} Samples were transferred in air to the nanoARPES end-station at beamline I05 at Diamond Light Source, United Kingdom. Prior to measurements the samples were annealed up to 450 $^{\circ}$C and kept under UHV conditions (pressure better than 10$^{-10}$~mbar) for the entire experiment. Synchrotron light with a photon energy of 95 eV was focused using a Fresnel zone plate followed by an order sorting aperture placed 8 and 4~mm from the sample, respectively. The sample was aligned using the characteristic linear dispersion of the underlying graphene substrate as described in Supplementary Section 1. The standard scanning mode involved collecting photoemission spectra with a Scienta Omicron DA30 hemispherical analyzer by rastering the sample position with respect to the focused synchrotron beam in steps of 250 nm using SmarAct piezo stages. Areas with WS$_2$ islands were found using coarse scan modes with larger step sizes as described in Supplementary Section 2. Typical data acquisition times for the scans presented here were on the order of  8 hours. The energy- and angular-resolution were set to 30 meV and 0.2$^{\circ}$, respectively. The spatial resolution was determined to be $(500 \pm 100)$ nm using a sharp feature in the sample as described in Supplementary Section 7. The experiments were carried out with the sample held at room temperature.

\section{acknowledgement}
We thank Diamond Light Source for access to Beamline I05 (Proposal No. SI19260) that contributed to the results presented here. S.U. acknowledges financial support from VILLUM FONDEN under the Young Investigator Program (grant no. 15375). This project has received funding from the European Union's Horizon 2020 research and innovation programme under grant agreement Graphene Core2 785219. C. E. G. acknowledges financial support from the UK National Measurement System.  
 J. A. M. and Ph. H. acknowledge support from the Danish Council for Independent Research, Natural Sciences under the Sapere Aude program (Grants No. DFF-6108-00409 and DFF  4002-00029) and the Aarhus University Research Foundation. This work was supported by VILLUM FONDEN via the Centre of Excellence for Dirac Materials (Grant No. 11744). D. K. G. and R. L. M. W. and work at NRL was supported by the Office of Naval Research. We thank Davide Curcio and Marco Bianchi for help with initial sample characterization. 
%
%\section{Author Contributions}
%contributions.
%
%

\section{Author information}
The authors declare that they have no competing financial interests.
Supplementary Information accompanies this paper.
Correspondence and requests for materials should be addressed to S. Ulstrup (ulstrup@phys.au.dk) or C. Giusca (cristina.giusca@npl.co.uk).

%\bibliography{refs}

\end{document}